\documentclass[12pt]{article}
\usepackage{graphicx,rotating,axodraw,caption2,color}
\usepackage{amsmath,amssymb,cite}
\usepackage[T1]{fontenc} % used for portuguese accents
\usepackage{cite}

\allowdisplaybreaks

\def\a{\alpha}
\def\b{\beta}

\def\f{\phi}

\def\j{\psi}

\def\m{\mu}
\def\n{\nu}

\def\p{\pi}

\def\t{\tau}

\def\D{\Delta}
\def\F{\Phi}

\def\L{\Lambda}

\def\Q{\Theta}

\def\tm{\widetilde{\mu}}

\def\tM{\widetilde{M}}

\def\NO{\nonumber}

\def\pl#1#2#3{Phys.~Lett.~{\bf B {#1}} ({#2}) #3}
\def\np#1#2#3{Nucl.~Phys.~{\bf B {#1}} ({#2}) #3}
\def\prl#1#2#3{Phys.~Rev.~Lett.~{\bf #1} ({#2}) #3}
\def\pr#1#2#3{Phys.~Rev.~{\bf D {#1}} ({#2}) #3}

\def\prep#1#2#3{Phys.~Rep.~{\bf {#1}C} ({#2}) #3}

\newcommand{\VEV}[1]{\langle #1 \rangle}

\renewcommand{\(}{\left(}
\renewcommand{\)}{\right)}
\renewcommand{\[}{\left[}

\newcommand{\tr}{\mbox{tr}}

\newcommand{\ckm}{\text{\sc ckm}}
\newcommand{\GG}{\text{\sc gg}}
\newcommand{\Gg}{\text{\bfseries\scshape gg}}

\newcommand{\ps}{\text{\sc ps}}
\newcommand{\Ps}{\text{\bfseries\scshape ps}}
\newcommand{\sm}{\text{\sc sm}}

\begin{document}
\date{\mbox{ }}
\title{
  {\normalsize 
                                %\hfill\mbox{}\\
    \hfill CERN-PH-TH/2004-043\mbox{}\\
    \hfill DESY 03-202\mbox{\hspace{2cm}}\\
    July 2004\hfill\mbox{}
    }\\
  \vspace{2cm}
  \textbf{Flavour Structure and Proton Decay\\
    in 6D Orbifold GUTs}\\
  [8mm]}

\author{W.~Buchm\"uller$^1$, L.~Covi$^2$, \\D.~Emmanuel-Costa$^3$, 
  S.~Wiesenfeldt$^1$\\
  \\
  {\normalsize
    $^1$ Deutsches Elektronen-Synchrotron DESY, Hamburg, Germany}
  \\
  {\normalsize
    $^2$ Theory Division, CERN Department of Physics, Geneva,
    Switzerland}
  \\
  {\normalsize  
    $^3$ CFTP, %Grupo de Física Teórica de Partículas, 
    Departamento de Física,  
    Instituto Superior Técnico, Lisbon, Portugal}}

\maketitle

\thispagestyle{empty}

\begin{abstract}
  \noindent
  We study proton decay in a supersymmetric {\sf SO(10)} gauge theory
  in six dimensions compactified on an orbifold.  The dimension-5
  proton decay operators are forbidden by R-symmetry, whereas the
  dimension-6 operators are enhanced due to the presence of KK towers.
  Three sequential quark-lepton families are localised at the three
  orbifold fixed points, where {\sf SO(10)} is broken to its three GUT
  subgroups.  The physical quarks and leptons are mixtures of these
  brane states and additional bulk zero modes.  This leads to a
  characteristic pattern of branching ratios in proton decay, in
  particular the suppression of the $p\rightarrow \m^+K^0$ mode.
\end{abstract}

\newpage

\section{Introduction \label{se:intro}}

The most striking consequence of Grand Unified Theories is proton
decay \cite{PS,GG,FM1,FM2,flipped1,flipped2}.  It has been predicted
and sought for more than 30 years, and its absence constitutes a very
strong constraint on any realistic GUT model.

Recently, the discussion of proton decay has been revitalised on two
different fronts.  On the experimental side, the bounds coming from
the SuperKamiokande experiment have reached \mbox{$\tau (p\rightarrow
  e^+\pi^0) \geq 5.3\times 10^{33}$~yrs} \cite{SK-epi} and \mbox{$\tau
  (p\rightarrow K^+ \bar \nu ) \geq 1.9\times 10^{33}$ yrs}
\cite{SK-Knu}.  On the theoretical side, these bounds have motivated
new detailed studies of proton decay via dimension-5 operators in
supersymmetric {\sf SU(5)} models \cite{gn99,dmr00,mp02,bfs02,cw03}.
These analyses showed that the present bounds disfavour this class of
models, although the theoretical predictions strongly depend on the
flavour structure assumed \cite{bfs02,cw03}.

Dimension-6 operators are less dangerous and have not drawn so much
attention in the literature.  It is, however, well known that such
operators depend on the flavour structure as well
\cite{jar79,moh79,egn79}, and it has been realised that the observed
leptonic mixing can have a strong effect on the proton decay branching
ratios \cite{ar01}.  Recently, dimension-6 operators have been studied
in flipped {\sf SU(5)} \cite{enw02} and {\sf SO(10)} \cite{ar01,fp04}
GUTs.

The interest in dimension-6 operators has been renewed by GUT models
in higher dimensions, with symmetry breaking via orbifolds.  Here
dimension-5 proton decay is naturally absent \cite{altarelli01,hn01}.
In such models though, the dimension-6 operators are enhanced compared
to the usual 4D case, due to the lower mass scale for the heavy
particles mediating the decay and the presence of Kaluza-Klein towers
of such states.  Furthermore, the proton decay branching ratios depend
on the localization of quarks and leptons~\cite{hm02}.

The goal of the present paper is to study proton decay via dimension-6
operators in a specific {\sf SO(10)} orbifold GUT model in 6D, where
the complete Standard Model flavour structure can be reproduced via
mixing of brane states with bulk split multiplets \cite{abc03b}.  This
kind of model can naturally arise in compactifications of the
heterotic string \cite{raby04,fnx04}.  Quark and lepton mass matrices
are approximately of lopsided type, with some characteristic
modifications compared to the 4D lopsided picture.  The up quark of
the first generation is located on a brane where the bulk gauge group
{\sf SO(10)} is broken to {\sf SU(5)xU(1)}, and therefore {\sf SU(5)}
gauge boson exchange gives the main contribution to proton decay.  For
this model, we will calculate the total rate and the branching ratios
of proton decay and compare the results with those in a 4D {\sf SU(5)}
models with {\sf U(1)} family symmetry.

The paper is organised as follows: in Section~\ref{se:model}, we
discuss the flavour structure of the 6D orbifold GUT model and
evaluate the mixing matrices needed to calculate proton decay widths.
Section~\ref{se:dim6} deals with proton decay via dimension-6
operators in the usual 4D case as well as the 6D model for which the
sum over the Kaluza-Klein tower is performed.  Finally, in
Section~\ref{se:br} we discuss the effects of the flavour structure on
the proton decay branching ratios in the different models.
Conclusions are given in Section~\ref{se:concl}.

\section{6D orbifold GUT model \label{se:model}}

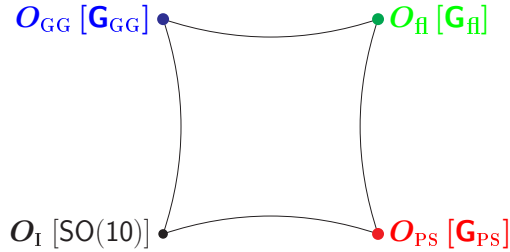
\begin{figure}[t]
  \centering
  \scalebox{.45}{
    \begin{picture}(200,200)(0,0)    
      \Curve{(10,10)(100,25)(190,10)}
      \Curve{(10,190)(100,175)(190,190)}
      \begin{sideways}
        \Curve{(10,-10)(100,-25)(190,-10)}
        \Curve{(10,-190)(100,-175)(190,-190)}
      \end{sideways}
      \Vertex(10,10)4  
      \Text(0,10)[r]{\huge{\boldmath{$O_\text{\bfseries\scshape i}$}}
        $\left[{\sf SO(10)}\right]$} 
      \SetColor{Blue}
      \Vertex(10,190)5 
      \Text(0,190)[r]{\huge\boldmath{${\color{blue}{O_\Gg \left[{\sf
                  G_\Gg}\right]}}$}}
      \SetColor{Green}
      \Vertex(190,190)5 
      \Text(200,190)[l]{\huge\boldmath{${\color{green}{O_\text{\bf fl}
              \left[{\sf G_\text{\bf fl}}\right]}}$}}
      \SetColor{Red}
      \Vertex(190,10)5
      \Text(200,10)[l]{\huge\boldmath{${\color{red}{O_\Ps \left[{\sf
                  G_\Ps}\right]}}$}}
    \end{picture}
    }
  \caption{The three {\sf SO(10)} subgroups at the corresponding fixed
    points (branes) of the orbifold \mbox{$T^2/Z_2 \times Z_2^\prime
      \times Z_2^{\prime\prime}$}.}
  \label{fig:orb}
\end{figure}

We start from an {\sf SO(10)} gauge theory in 6D with $N=1$ %SUSY
supersymmetry 
compactified on the orbifold $T^2/(Z^I_2\times Z^{\ps}_2\times
Z^{\GG}_2)$ \cite{abc01,hnx02}. The theory has four fixed points,
$O_\text{\sc i}$, $O_{\ps}$, $O_{\GG}$ and $O_\text{fl}$, located at
the four corners of a `pillow' corresponding to the two compact
dimensions (cf. Fig. \ref{fig:orb}).  At $O_\text{\sc i}$, only SUSY
%supersymmetry 
is broken, whereas at the other fixed points, $O_{\ps}$,
$O_{\GG}$ and $O_\text{fl}$, also %the gauge group 
{\sf SO(10)} is broken to its three GUT 
subgroups \mbox{${\sf G}_{\ps}={\sf SU(4)}\times {\sf SU(2)} \times
  {\sf SU(2)}$} \cite{PS}, ${\sf G}_{\GG}={\sf SU(5)}\times {\sf
  U(1)}_X$ \cite{GG} and flipped {\sf SU(5)} \cite{flipped1,flipped2},
\mbox{${\sf G}_\text{fl}={\sf SU(5)'}\times {\sf U(1)'}$},
respectively. The intersection of all these GUT groups yields the
standard model group with an additional {\sf U(1)} factor, \mbox{${\sf
    G}_{\sm '}= {\sf SU(3)}\times {\sf SU(2)} \times {\sf U(1)}_Y
  \times {\sf U(1)}_{Y'}$}, as unbroken gauge symmetry below the
compactification scale.

The field content of the theory is strongly constrained by imposing
the cancellation of irreducible bulk and brane anomalies
\cite{ss04,abc03}.  We study the model proposed in \cite{abc03b},
containing three {\bf 16}-plets $\j_i$, $i=1,\ldots,3$, as brane
fields, and six {\bf 10}-plets, $H_1,\ldots, H_6$, and four {\bf
  16}-plets, $\F, \F^c, \f, \f^c$, as bulk hypermultiplets.  Vacuum
expectation values of $\F$ and $\F^c$ break the surviving ${\sf
  U(1)}_{B-L}$.  The electroweak gauge group is broken by expectation
values of the anti-doublet and doublet contained in $H_1$ and $H_2$.
 
%------------------------Figure---------------------------------------
%--------------------End Figure---------------------------------------

We choose the parities of $H_5$, $H_6$ and $\f,\f^c $ such that their
zero modes are 
\begin{align}
  L = \left( \begin{array}{l} 
      \n_4 \\ e_4
    \end{array} \right)\;, \quad  L^c = \left( \begin{array}{l} 
      \n^c_4 \\ e^c_4 
    \end{array}\right)\;, \quad G^c_5 = d^c_4\;, \quad G_6 = d_4\;.
\end{align}
These zero modes act as a (vectorial) fourth generation of down quarks
and leptons and mix with the three generations of brane fields.  We
allocate the three sequential {\bf 16}-plets to the three branes where
{\sf SO(10)} is broken to its three GUT subgroups, and place $\j_1$ at
$O_{\GG}$, $\j_2$ at $O_\text{fl}$ and $\j_3$ at $O_{\ps}$.  The three
`families' are then separated by distances large compared to the
cutoff scale $M_*$. Hence, they can only have diagonal Yukawa
couplings with the bulk Higgs fields. Direct mixings are exponentially
suppressed.  The brane fields, however, can mix with the bulk zero
modes for which we expect no suppression. These mixings take place
only among left-handed leptons and right-handed down-quarks.  This
leads to a characteristic pattern of mass matrices.

As described in \cite{abc03b}, after $B-L$ breaking at the scale $v_N$
and electroweak symmetry breaking via $v_1 = \langle H^c_1 \rangle$,
$v_2 = \langle H_2 \rangle$, the mass terms assume the characteristic
form
\begin{align}
  W & = d_\a m^d_{\a\b} d^c_\b + e^c_\a m^e_{\a\b} e_\b + n^c_\a
  m^D_{\a\b} \n_\b + u_i m^u_i u^c_i + \frac{1}{2} n^c_i m^N_i n^c_i\;,
\end{align}
where the Greek indices, $\alpha = 1\ldots 4$, include also the bulk
states with $n^c_4 = \nu^c_4$, while Latin indices only run over brane
fields, $i=1\ldots 3$. Note that there is no Majorana mass for $n_4^c
= \nu_4^c$ which originates from the 6D hypermultiplet.  $m^u$ and
$m^N$ are diagonal $3\times 3$ matrices ($\tan{\beta}=v_2/v_1$),
\begin{align} \label{muN}
  \frac{1}{\tan{\beta}}\, m^u \sim \frac{v_1 M_*}{v_N^2}\ m^N \sim 
  \left(\begin{array}{ccc}
      \mu_1 & 0 & 0 \\
      0 & \mu_2 & 0 \\
      0 & 0 & \mu_3 \end{array}\right) ,
\end{align}
whereas $m^d$, $m^e$ and $m^D$ are $4\times 4$ matrices with the
common structure,
\begin{align} \label{eq:matrix-structure}
  \frac{1}{\tan{\beta}}\ m^D \sim m^d \sim m^e \sim
  \left(\begin{array}{cccc}
      \m_1 & 0 & 0 & \tm_1 \\
      0 & \m_2 & 0 & \tm_2 \\
      0 & 0 & \m_3 & \tm_3 \\
      \tM_1 & \tM_2 & \tM_3 & \tM_4
    \end{array} \right) \equiv m \ .
\end{align}
Here $\m_i,\tm_i = {\cal O}(v_{1})$ and $\tM_i = {\cal O}(\L_{GUT})$.
The diagonal and the off-diagonal elements of these matrices satisfy
several relations due to the underlying GUT symmetries \cite{abc03b}.
The hypothesis of a universal strength of Yukawa couplings at each
fixpoint leads to the identification of the diagonal and off-diagonal
elements of $m^u/\tan{\beta}$, $m^d$, $m^e$ and $m^D/\tan{\beta}$ up
to coefficients of order one.  This implies an approximate top-bottom
unification with large $\tan{\beta}$ and a parametrization of quark
and lepton mass hierarchies in terms of the six parameters
$\m_1,\m_2,\m_3$ and $\tm_1,\tm_2,\tm_3$.

The crucial feature of the matrices $m^d$, $m^e$ and $m^D$ are the
mixings between the six brane states and the two bulk states. The
first three rows of the matrices are proportional to the electroweak
scale.  The corresponding Yukawa couplings have to be hierarchical in
order to obtain a realistic spectrum of quark and lepton masses.  In a
complete theory, this hierarchy may be due to the different location
of the fixpoints on the orbifold. The fourth row, $\tM_{\alpha}$, is
of order the unification scale and, as we assume, non-hierarchical.

The mass matrix $m$ (cf.~Eq.~(\ref{eq:matrix-structure})) can be
diagonalised by the unitary transformation
\begin{align}
  m = U_4 U_3 D V_3^\dagger V_4^\dagger \;,
\end{align}
where the matrices $U_4,V_4$ single out the heavy mass eigenstate,
that can then be integrated out, while the matrices $U_3,V_3$ act only
on the SM flavour indices and perform the final diagonalisation in the
remaining $3\times 3$ subspace.  $U_4$ and $V_4$ are given by
(neglecting phases)
\begin{align}
  U_4 &= \left(\begin{array}{cccc} 1&0&0&
      \frac{\mu_1\widetilde{M}_1 +
        \tm_1\widetilde{M}_4}{\widetilde{M}^2} \\
      \rule[-3mm]{0mm}{9.5mm}
      0&1&0& \frac{\mu_2\widetilde{M}_2 +
        \tm_2\widetilde{M}_4}{\tilde
        M^2} \\
      \rule[-3mm]{0mm}{9mm}
      0&0&1& \frac{\mu_3\widetilde{M}_3 +
        \tm_3\widetilde{M}_4}{\tilde
        M^2} \\
      \rule[0mm]{0mm}{3mm}
      -\frac{\mu_1 \widetilde{M}_1 + \tm_1
        \widetilde{M}_4}{\widetilde{M}^2} & -\frac{\mu_2
        \widetilde{M}_2 + \tm_2
        \widetilde{M}_4}{\widetilde{M}^2} & -\frac{\mu_3
        \widetilde{M}_3 + \tm_3
        \widetilde{M}_4}{\widetilde{M}^2}& 1
    \end{array}\right) + {\cal O}\left(\frac{v^2}{\tM^2}\right),
  \\[2mm]
  V_4 &= \left(\begin{array}{cccc}
      \frac{\widetilde{M}_4}{\widetilde{M}_{14}} & 0 &
      -\frac{\widetilde{M}_1\,
        \widetilde{M}_{23}}{\widetilde{M}\,\widetilde{M}_{14}} &
      \frac{\widetilde{M}_1}{\widetilde{M}}
      \\
      \rule[-2mm]{0mm}{8.5mm} 0 &
      \frac{\widetilde{M}_3}{\widetilde{M}_{23}} &
      \frac{\widetilde{M}_2\, \widetilde
        M_{14}}{\widetilde{M}\,\widetilde{M}_{23}} &
      \frac{\widetilde{M}_2}{\widetilde{M}}
      \\
      \rule[-2mm]{0mm}{9mm} 0 &
      -\frac{\widetilde{M}_2}{\widetilde{M}_{23}} &
      \frac{\widetilde{M}_3\,
        \widetilde{M}_{14}}{\widetilde{M}\,\widetilde{M}_{23}} &
      \frac{\widetilde M_3}{\widetilde{M}}
      \\
      \rule[0mm]{0mm}{6mm} -\frac{\widetilde{M}_1}{\widetilde{M}_{14}}
      & 0 & -\frac{\widetilde{M}_4\,
        \widetilde{M}_{23}}{\widetilde{M}\,\widetilde{M}_{14}} &
      \frac{\widetilde M_4}{\widetilde{M}}
    \end{array}\right)  ,
\end{align}
where $\tM = \sqrt{\sum_\alpha \tM_\alpha^2}$ and
$\tM_{\alpha\beta}=\sqrt{\tM_\alpha^2+\tM_\beta^2}$.  In general,
$V_4$ contains large mixings, while $U_4$ is approximately the unit
matrix up to corrections ${\cal O}(v/\tM)$. $ U_3$ and $V_3$ are the
matrices that diagonalise
\begin{align}
  m' = U_4^{\dagger} m V_4 = \left(
    \begin{array}{cc}
      \widehat{m} & 0 \\ 0 & \tM
    \end{array}\right) + {\cal O}\left(\frac{v^2}{\tM}\right),
\end{align}
where
\begin{align} \label{mass3}
  \widehat{m} = 
  \left( \begin{array}{c}
      \mu_1 (V_4)_{1j} + \tm_1 (V_4)_{4j}\\
      \mu_2 (V_4)_{2j} + \tm_2 (V_4)_{4j}\\
      \mu_3 (V_4)_{3j} + \tm_3 (V_4)_{4j}
    \end{array} \right) .
\end{align}
Clearly, they have only a non-trivial $3\times 3$ part,
\begin{align}
  U_3 & = \left(\begin{array}{cc}
      V_\ckm^\dagger & 0\\
      0 & 1
    \end{array}\right) , &
  V_3 & = \left(\begin{array}{cc}
      \widehat V  & 0 \\
      0  & 1 
    \end{array}\right) .
\end{align}
Notice that the rows of $\widehat{m}$ scale each like $\mu_i , \tm_i$.
Hence, for hierarchical parameters $\mu_1,\tm_1 < \mu_2,\tm_2 < \mu_3,
\tm_3$ we obtain a structure familiar from lopsided fermion mass
models.

As discussed in \cite{abc03b}, we can choose the parameters in such a
way to give a consistent quark mass pattern and CKM matrix, in
particular
\begin{equation} \label{diag} 
  \m_1 : \m_2 : \m_3  \sim m_u : m_c : m_t \; , 
\end{equation} 
and (cf.~Eq.~(\ref{eq:b3})) 
\begin{equation} 
  \bar\m_3  \simeq m_b \; , \quad
  \tm_2 : \bar\m_3  \sim m_s : m_b \; .
\end{equation}
The CKM matrix in $U_3$ arises in a natural way as well. Setting the
remaining parameter $\tm_1 $ to give
\begin{equation}
  V_{us} = \Q_c \simeq \frac{\tm_1}{\tm_2} \simeq 0.2\;, \quad
\end{equation}
the other matrix elements are determined by the quark masses
\cite{abc03b},
\begin{align}
  V_{cb} & \sim \frac{\tm_2}{ \tm_3} \simeq 
  \frac{m_s}{ m_b} \simeq 2\times 10^{-2} \; , &
  V_{ub} & \sim \frac{\tm_1 }{\tm_3} = 
  \Q_c \frac{m_s}{ m_b} \simeq 4\times 10^{-3} \; .
\end{align}
Within the accuracy of our approach this is consistent with the
analysis from weak decays \cite{fle02}.

The matrix $\widehat V$ takes a relatively simple form in the case
when $\mu_{1,2} = 0$ and the mass matrix in Eq.~(\ref{mass3}) has one
zero eigenvalue.  We then obtain
\begin{align} 
  \widehat V = \left( \begin{array}{ccc} -\frac{\tM_2\,
        \tM_4}{\tM_{12}\, \tM_{14}} & \frac{\tM_1 \left( \tm_3
          \tM_3\, \tM_4-\mu_3 \left( \tM_1^2+\tM_2^2+\tM_4^2 \right)
        \right)}{\bar\mu_3\, \tM\, \tM_{12}\, \tM_{14}} &
      -\frac{\tm_3}{\bar\mu_3}\frac{\tM_1}{\tM_{14}}
      \\
      \rule[-5mm]{0mm}{13mm} \frac{\tM_1\, \tM_3}{\tM_{12}\, \tM_{23}}
      & \frac{\tM_2 \left( \tm_3 \left( \tM_1^2+\tM_2^2+\tM_3^2
          \right) -\mu_3\tM_3\, \tM_4 \right)}{\bar\mu_3\, \tM\,
        \tM_{12}\, \tM_{23}} &
      -\frac{\mu_3}{\bar\mu_3}\frac{\tM_2}{\tM_{23}}
      \\
      \rule[-5mm]{0mm}{13mm} \frac{\tM_1\, \tM_2\, \tM}{\tM_{12}\,
        \tM_{14}\, \tM_{23}} & -\frac{\tm_3 \tM_1^2\,
        \tM_3+\mu_3\tM_2^2\, \tM_4}{\bar\mu_3\, \tM_{12}\, \tM_{14}\,
        \tM_{23}} & -\frac{\tm_3}{\bar\mu_3}\frac{\tM_4\,
        \tM_{23}}{\tM\, \tM_{14}} + \frac{\mu_3}{\bar\mu_3}
      \frac{\tM_3\, \tM_{14}}{\tM\, \tM_{23}}
    \end{array} \right),
  \label{eq:vhat}
\end{align}
up to a two-dimensional mixing matrix for the second and third
generation, parameterized by a small angle $\Theta_R$,
\begin{align}
  \Theta_R \simeq \frac{\tm_1^2+\tm_2^2}{\bar\mu_3^2} \ll 1 \;,
\end{align}
where we have defined
\begin{align} \label{eq:b3}
  \bar\mu^2_3 & = \tm_3^2 \left( 1- \frac{\tM_4^2}{\tM^2}\right)
  + \mu_3^2 \left(1- \frac{\tM_3^2}{\tM^2}\right) - 2\mu_3\tm_3
  \frac{\tM_3\tM_4}{\tM^2} \ .
\end{align}

The case of small $\mu_{1,2} $ limit is actually of physical
relevance, since it gives for the down quark
\begin{align}\label{md}
  \frac{m_d}{ m_s} \sim \frac{\m_2}{ \tm_2}\,\frac{\tm_1}{\tm_2} \sim
  \ \Q_c \; \frac{m_c m_b}{ m_t m_s} \simeq\ 0.03\;,
\end{align}
consistent with data \cite{fx00}.  Since $\m_2/\tm_2 \sim m_c m_b/(m_t
m_s) \sim 0.1$, the corrections to the matrix (\ref{eq:vhat}) are
small.

The charged lepton mass matrix $m^e$ has the same structure as the
down-quark mass matrix, but there the large mixings are between the
`left-handed' states $e_i$.  Experimental data require for the largest
eigenvalue $m_\tau \simeq m_b$, whereas the second and the third
eigenvalue have to satisfy the relations $m_\m \simeq 3 m_s$ and $m_e
\simeq 0.2 m_d$, respectively.  This is consistent with our
identification of $m^d$ and $m^e$ up to coefficients of order one
unless the relevant parameters are fixed by GUT relations.  The
comparision between the expression (\ref{md}) for $m_d$ and the
corresponding one for $m_e$ suggests locating the second family on the
flipped {\sf SU(5)} brane.  The successful relation for the light
neutrino mass, \mbox{$m_3 \sim m_t^2/M_3 \sim m_t^2 M_*/v_N^2 \sim 0.01$~eV},
requires the third family to be located on the PS brane.  With the
first family on the GG brane, $m_d$ and $m_e$ are determined by the
parameters $\m_2^d$ and $\m_2^e$, which are not related by a flipped
{\sf SU(5)} mass relation.

\section{Proton decay via dimension-6 operators \label{se:dim6}}

\subsection{Effective {\sf SU(5)} operators in 4D models}

Dimension-6 proton decay in the {\sf SU(5)} model is mediated by the
exchange of the $X$ and $Y$ leptoquark gauge bosons \cite{egn79}.
Contrary to the dimension-5 operator, it does not involve any dressing
through supersymmetric partners and therefore it is not sensitive to
the supersymmetry breaking scale (except for the weak dependence of
the GUT scale on the superparticle mass spectrum). The effective
vertex is obtained by simply integrating out the heavy gauge bosons.

The couplings of the {\sf SU(5)} representations ${\bf 5^*} $ and
${\bf 10}$ with the {\sf SU(5)} gauge bosons are given by their
kinetic terms,
\begin{align}
  \int_{\theta^2 \bar\theta^2} \sum_\text{reps} \bar \Phi_i e^{2V}
  \Phi_i \ ,
\end{align}
which include 
\begin{align}
  {\cal L} = i \frac{g_5 }{ \sqrt{2}}\; A_\mu^a \left[ 2\, \tr
    \left({\bf \overline{10}_i}\, \gamma^\mu T^a\, {\bf 10_i}\right) +
    {\bf \bar 5^*_k}\, \gamma^\mu (T^a)^\top\, {\bf 5^*_k} \right] +
  \text{h.c.}\;,
\end{align}
where $g_5$ is the {\sf SU(5)} gauge coupling; the Latin indices count
the generations, $i=1 \ldots 3 $ those of ${\bf 10}$, $k=1 \ldots 3$
those of ${\bf 5^*}$.

We now express the {\sf SU(5)} representations in terms of SM fields
with ${\cal X} = (X,Y)$ being a $({\bf 3^*},{\bf 2},5/6)$
representation of \mbox{${\sf SU(3)}\times{\sf SU(2)}\times{\sf
    U(1)}$}.  This yields the baryon and lepton number violating terms
\begin{align}
  {\cal L} = -i \frac{g_5}{ \sqrt{2}} {\cal X}_{\alpha\mu} \left[
    \epsilon_{\alpha\beta\gamma}\, \overline{Q}_{\beta,i}\,
    \gamma^\mu\, u^c_{\gamma,i} + \overline{e^c}_i\, \gamma^\mu\,
    Q_{\alpha,i} - \overline{d^c}_{\alpha,k}\, \gamma^\mu\, L_k
  \right] + \text{h.c.}\;,
\end{align}
where Greek indices denote the colour degrees of freedom and the {\sf
  SU(2)} indices have been suppressed.  Note that the first two terms
come from the $ {\bf 10}$ representation, the last one from the ${\bf
  5^*}$.

Integrating out the heavy gauge bosons with masses $M_{\cal X}$, we
get the effective operators relevant for proton decay
\begin{align}
  {\cal L}_\text{eff} &= - \frac{g_5^2 }{ 2 M_{\cal X}^2}\;
  \epsilon_{\alpha\beta\gamma}\;\overline{u^c}_{\!\!\alpha, i}\,
  \gamma^\mu\, Q_{\beta,i}\; \left[ \overline{e^c}_{\!\!j}\,
    \gamma_\mu\, Q_{\gamma,j}\, - \overline{d^c}_{\!\!\gamma,k}\,
    \gamma_\mu\, L_k \right] + \text{h.c}. \; .
  \label{eff4D}
\end{align}
With Fierz reordering, one can write the operators as
\begin{align}
  {\cal L}_\text{eff} &= - \frac{g_5^2 }{ M_{\cal X}^2}\;
  \epsilon_{\alpha\beta\gamma} \left[ \overline{e^c}_{\!j}
    \overline{u^c}_{\!\alpha,i}\, Q_{\beta,i}\, Q_{\gamma,j} -
    \overline{d^c}_{\!\alpha,k} \overline{u^c}_{\!\beta,i}\,
    Q_{\gamma,i}\, L_k \right] + \text{h.c.}  \, .
  \label{effWeyl-RRLL}
\end{align}

In the case of flipped {\sf SU(5)} the ${\cal X'} = (X',Y')$ bosons
form a $({\bf 3^*},{\bf 2},-1/6)$ representation of \mbox{${\sf
    SU(3)}\times{\sf SU(2)}\times{\sf U(1)}$} with couplings
\begin{align}
  {\cal L} = -i \frac{g_5}{ \sqrt{2}} {\cal X}^\prime_{\alpha\mu}
  \left[ 
    \epsilon_{\alpha\beta\gamma}\, \overline{Q}_\beta\, \gamma^\mu\,
    d^c_\gamma - \overline{u^c}_\alpha\, \gamma^\mu\, L 
  \right] + \text{h.c.}\;.
\end{align}
Contrary to {\sf SU(5)}, there is only a single baryon and lepton
number violating operator in flipped {\sf SU(5)},
\begin{align}
  {\cal L}_\text{eff} = \frac{g_5^2 }{ M_{\cal X}^{\prime\,2}}\;
  \epsilon_{\alpha\beta\gamma} 
    \overline{d^c}_{\!\alpha,k} \overline{u^c}_{\!\beta,i}\,
    Q_{\gamma,i}\, L_k  + \text{h.c.}  \, .
\end{align}

\subsection{Effective operators in 6D models}

In the orbifold model described above, the up-type quarks are
localized at one fixed point each, in particular the up quark is
located at the Georgi-Glashow one. It is therefore clear that
dimension-6 proton decay can arise via the exchange of the {\sf SU(5)}
$X$ and $Y$ bosons as in the traditional 4D picture.  There are though
two important differences in the 6D case, as we will see in the
following.

First we have to take into account the presence of not only one ${\cal
  X}$ gauge boson, but of a Kaluza-Klein (KK) tower with masses given
by
\begin{align}
  M_{\cal X}^2 (n,m) = \frac{(2n+1)^2}{R_5^2} + \frac{(2m)^2}{R_6^2}
  \;, 
\end{align}
for $n,m = 0,1,2, \ldots $.  The lowest possible mass is $M_{\cal
  X}(0,0)=1/R_5$, as given by the {\sf SU(5)} breaking parity.  Note
that if we define the 4D gauge coupling as the effective coupling of
the zero modes, the KK modes interact more strongly by a factor
$\sqrt{2} $ due to their bulk normalization.

To obtain the low energy effective operator, we then have to sum over
the KK modes,
\begin{align}
  \frac{1}{(M_{\cal X}^\text{eff})^2}
  & = 2 \sum_{n,m=0}^\infty \frac{1}{ M_{\cal X}^2 (n,m)} \NO\\
  & = 2 \sum_{n,m=0}^{\infty} \frac{R_5^2}{(2 n +1)^2 +
    \frac{R_5^2}{R_6^2} (2m)^2}\;.
  \label{sumKK}
\end{align}
Taking formally the limit $R_6/R_5 \rightarrow 0$, we regain the 5D
result \cite{hm02},
\begin{align}
  \frac{1}{ (M_{\cal X}^\text{eff})^2} = 2\sum_{n=0}^{\infty}
  \frac{R_5^2}{(2n+1)^2} = \frac{\pi^2 R_5^2}{4}\;.
  \label{eq:coupling-constant}
\end{align}

The double sum in Eq.~(\ref{sumKK}) is logarithmically divergent.
Since our model is valid only below the scale $M_*$, where it becomes
strongly coupled and also 6D gravity corrections are no longer
negligible, we restrict the sum to masses $M_{\cal X}(n,m)\leq M_*$.
One easily finds
\begin{align}\label{sumKKM}
  \frac{1}{(M_{\cal X}^\text{eff})^2} = \frac{\pi}{4}\, R_5 R_6
  \left(\ln \left(M_* R_5 \right) + C\left(\frac{R_5}{R_6}\right) +
    {\cal O}\left(\frac{1}{R_{5/6}\, M_*}\right)\right) \;.
\end{align}
Note that in the logarithm the smallest KK mass appears,
$R_5=1/M_{\cal X}(0,0)$.  The dependence of $1/(M_{\cal
  X}^\text{eff})^2$ on the cutoff $M_*$ has to disappear once the
model is embedded in a more fundamental theory.  In the symmetric
case, $R_5 = R_6 = 1/M_c$, one finds $C(1) \simeq 2.3$ and the
expression (\ref{sumKKM}) simplifies to
\begin{align}\label{can}
  \frac{1}{(M_{\cal X}^\text{eff})^2} \simeq \frac{\pi}{
    4\,M_c^2}\(\ln \(\frac{M_*}{M_c}\) + 2.3 \) \; .
\end{align}
Numerically, this agrees with the explicit sum over the KK masses
within 1\% for $M_*/M_c = 10\ldots 50$, which is the relevant range
for the ratio of cutoff and compactification scales in 6D.

The second, most important difference of 6D models compared to 4D
models is the non-universal coupling of the ${\cal X}$ gauge bosons.
In fact, due to the parities and the {\sf SO(10)} breaking pattern,
their wavefunctions must vanish on the fixed points with broken {\sf
  SU(5)} symmetry, $O_{\ps}$ and $O_\text{fl}$, and therefore no
coupling arises via the kinetic term with the charm and top quark or
to the brane states $d^c_{2,3}, l_{2,3}$.  We also have couplings to
the bulk states $d^c_4,d_4$ and $l_4, l^c_4$.  However, due to the
embedding of the zero modes in full {\sf SU(5)} multiplets together
with massive KK modes, i.e.  $(d^c_4, L_4)$, $(d_4, L^c_4)$, $(D^c_4,
l_4)$ and $(D_4, l^c_4)$, the charged current interaction always mixes
the light states with the heavy ones, and it is therefore irrelevant
for the low energy process of proton decay \cite{hm02}.  So the
kinetic coupling in Eq.~(\ref{effWeyl-RRLL}) arises only for a single
flavour eigenstate, not for all flavours as in the usual 4D case.

\subsection{Corrections from derivative brane operators}

Apart from the kinetic term couplings, at any brane additional
couplings can arise containing derivatives along the extra dimensions
of the locally vanishing gauge bosons. Such operators are a 6D
generalization of the 5D derivative operators discussed in
\cite{heb02,hm02},
\begin{align}
  {\cal L}_d = \sum_\text{fixed points} \delta_i \(z\)
  \frac{c^i_{5/6}}{M_*} \int_{\theta^2 \bar\theta^2} \bar \Phi_1 \;
  \({\cal D}_{5/6} e^{2V}\) \Phi_2 + \text{h.c.}\;.
  \label{eq:deriv-oper}  
\end{align} 
Here $c^i_{5/6} $ are unknown brane coefficients, ${\cal D}_{5/6} =
\partial_{5/6} + i A_{5/6} $ are the covariant derivatives in the
extra dimensions and $\Phi_i$ are any two different, locally
non-vanishing fields in group representations which form a singlet
together with generators of the broken symmetries.  For unbroken
symmetries the chiral superfields $A_{5/6}$ vanish at the fixpoints.

These supersymmetric terms produce on the GG brane couplings with the
flipped {\sf SU(5)} leptoquark gauge bosons ${\cal X'}$, whose
derivatives do not vanish on that brane.  On the flipped {\sf SU(5)}
brane, there are couplings containing the derivative of the ${\cal X}$
gauge bosons, and on the Pati-Salam brane there are derivative
couplings with both ${\cal X'}$ and ${\cal X}$.  Due to these
additional vertices, three different classes of operators can arise:
\begin{itemize} 
\item{operators coming from ${\cal X'}$ exchange on the GG brane:
    these involve two derivative vertices and can produce additional
    contributions to the effective operator
    \begin{align}
      \overline{d^c_k}\, \overline{u^c_1}\, Q_1 L_j
    \end{align} 
    with $k,j=1,4$; we will discuss their contribution below; }
\item{operators coming from ${\cal X}$-exchange on the flipped {\sf
      SU(5)} brane or ${\cal X},{\cal X}'$ exchange on the Pati-Salam
    brane: these usually involve the charm and top quark instead of
    the up quark, and they are therefore irrelevant for proton decay;
    }
\item{interbrane operators from both the exchanges of ${\cal X}$ and
    ${\cal X'}$ gauge bosons: they can involve either one or two
    derivative vertices and generate mixed flavour operators 
    of the type
    \begin{align}
      \overline{e^c_j}\, \overline{u^c_1}\, u_1\, d_k & -
      \overline{d^c_k}\, \overline{u^c_1}\, Q_1 L_j 
      & k,j & =2,3,4 & & \mbox{${\cal X}$ exchange,} 
      \\
      & \overline{d^c_2}\, \overline{u^c_1}\, d_2\, \nu_k &
      k & = 1,4 & & \mbox{${\cal X'}$ exchange GG---fl,}
      \\ 
      & \overline{d^c_j}\, \overline{u^c_1}\, d_l\, \nu_k &
      j,l & =3,4\,;\ k=1,4 & & \mbox{${\cal X'}$ exchange GG---PS.} 
    \end{align} 
    Apart from the last term, they are usually suppressed compared to
    the kinetic term operators by a factor $M_c/M_*$ due to the
    different parities of the vertices.}
\end{itemize}

To estimate the effect of these additional operators, which introduce
a dependence on the ${\cal O}(1)$ coefficients $c_{5/6}$, let us now
consider the KK summations with one or two derivative vertices.  We
restrict ourselves here to the case \mbox{$R_5 = R_6 = 1/M_c$} and
universal coefficients $c_{5/6}$ at the different fixpoints.  Note
that even if suppressed by $M_*$, these operators can be as important
as the usual ones, since the derivative enhances the divergence of the
KK summation, which compensates the suppression.  For example, from
the exchange of the ${\cal X'}$ bosons on the GG brane, we obtain the
sum
\begin{align}
  \frac{1}{(M_{\cal X}^\text{eff})^2_\text{b.o.}} = \frac{2}{M_*^2}
  \sum_{n,m} \frac{| c_5 \left(2n+1\right) + c_6 \left(2m+1\right)|^2}
  {\left(2n+1\right)^2 + \left(2m+1\right)^2}\; ,
\end{align} 
which is quadratically divergent. Using again the cutoff $M_{\cal
  X}(n,m)\leq M_*$, we obtain
\begin{align} \label{eq:sum-der}
  \frac{1}{(M_{\cal X}^\text{eff})^2_\text{b.o.}} = \frac{\pi}{16
    M_c^2} \left(|c_5|^2+ |c_6|^2 +\frac{4}{\pi} Re[c_5 c_6^*] + {\cal
      O}\left(\frac{M_c}{M_*}\right)\right) ,
\end{align} 
to be compared with the `canonical' term Eq.~(\ref{can}).  A similar
result is obtained for the exchange of ${\cal X}' $ gauge bosons
between the GG and the PS brane, affecting only the decay into
neutrinos.

The exchange of gauge bosons between different branes involving only
one single derivative operator are less dangerous since the propagator
gives a factor $(-1)^n $ accounting for the different parities of the
derivative and the kinetic term vertices.  Therefore the KK summations
are in general suppressed by a factor $M_c/M_*$.  The exchange of
${\cal X}$ bosons between the GG and the flipped {\sf SU(5)} brane
gives
\begin{align}
  \frac{1}{(M_{\cal X}^\text{eff})^2_\text{b.o.}} = \frac{2}{M_c M_*}
  \sum_{n,m} (-1)^n\, \frac{c_5 \left(2n+1\right) + c_6
    \left(2m\right)} {\left(2n+1\right)^2 + \left(2m\right)^2} \ .
\end{align}
This sum is only logarithmically divergent thanks to the alternating
signs. Using again the cutoff $M_{\cal X}(n,m)\leq M_*$, one finds
\begin{align} \label{eq:der-int}
  \frac{1}{(M_{\cal X}^\text{eff})^2_\text{b.o.}} = \frac{1}{2 M_c
    M_*} \left( c_6 \ln\left(\frac{M_*}{M_c}\right) + {\cal O}(1)
  \right) ,
\end{align}
which is suppressed compared to the `canonical' term by a factor
$M_c/M_*$. The same result is obtained for the ${\cal X'}$ exchange
between the GG and the flipped {\sf SU(5)} branes.

Note that in principle operators with a higher number of derivatives
can also be present, which contribute at the same level as the single
derivative ones because the divergence of the KK summation compensates
the suppression by powers of $M_*$.  We will assume that their
contribution is small.

Regarding the N=2 scalar superpartners of the {\sf SU(5)} gauge
bosons, ${\cal X}_{5,6}$, the brane terms in Eq.~(\ref{eq:deriv-oper})
give rise to couplings with fermion kinetic terms which can only
produce corrections of order $(m_p/M_*)^2$.  The derivatives of the
${\cal X}_{5,6}$ bosons do not couple to fermion pairs and are
therefore irrelevant.

Finally, we emphasize that the position of the lightest quark
generation is crucial for the discussion of proton decay. For
instance, if the up quark were located on the Pati-Salam brane, the
dimension-6 operator coming from the kinetic terms would be absent
since both the ${\cal X}$ bosons of {\sf SU(5)} and the ${\cal X}'$
bosons of flipped {\sf SU(5)} vanish there. In principle, this gives
us a means to avoid the `canonical' dimension-6 operators completely,
leaving the derivative couplings as dominant contributions.

\section{Flavour structure and branching ratios \label{se:br}}

\subsection{Flavour mixing in 6D versus 4D GUT models}

Proton decay involves only the light quark states and the operators
containing the combinations $uud$ and $udd$.  Therefore we have to
rotate the weak eigenstates into the mass eigenstates and single out
the contributions for the lightest generation.  Without loss of
generality and for future convenience, we can start in the basis where
the up-quark mass matrix is diagonal.  Then the down quark and lepton
mass matrices are not diagonal, in general, but can be diagonalised by
unitary transformations,
\begin{align}
  d_L & = U^d_L d'_L\;, & e_L & = U^e_L e'_L\;, & \nu_L & = U^{\nu}_L
  \nu'_L \;, \\
  d_R & = U^d_R d'_R\;, & e_R & = U^e_R e'_R\;,
\end{align}
where the prime denotes mass eigenstates. Since the up-quark matrix is
diagonal, $U^d_L$ coincides with the CKM matrix.

We can now express the proton decay operators of
Eq.~(\ref{effWeyl-RRLL}) in term of mass eigenstates,
\begin{multline}   \label{eff4D-VU}
  {\cal L}_\text{eff} = \frac{g_5^2}{ M_{\cal X}^2}\;
  \epsilon_{\alpha\beta\gamma} \left[ \overline{e^c}'_{\!\!k} \left(
      U^{e\top}_R \right)_{kj} \overline{u^c}_{\!\!\alpha,i} \left(
      d'_{\beta,m} \left( U^d_L \right)_{im} u_{\gamma,j} -
      u_{\beta,i} \left(U^d_L \right)_{jl} d'_{\gamma,l}
      \vphantom{e'_j} \right) \right.
  \\
  + \left. \overline{d^c}'_{\!\!\alpha,l} \left( U^{d\top}_R
    \right)_{lk} \overline{u^c}_{\!\beta,i} \left( u_{\gamma,i} \left(
        U^e_L \right)_{kj} e'_j - d'_{\gamma,m} \left( U^d_L
      \right)_{im} \left(U^{\nu}_L \right)_{kj} \nu'_j \right) \right]
  + \text{h.c.}
\end{multline}
Note that for the orbifold construction, where only the first
generation weak eigenstates couple to the ${\cal X}$ bosons, the
effective operators read instead
\begin{multline}  \label{eff6D-VU}
  {\cal L}_\text{eff} = \frac{g_5^2}{ (M_{\cal X}^\text{eff})^2}\;
  \epsilon_{\alpha\beta\gamma} \left[ 2\, \overline{e^c}'_{\!\!k}
    \left( U^{e\top}_R \right)_{k1} \overline{u^c}_{\!\!\alpha,1} \,
    d'_{\beta,m} \left( U^d_L \right)_{1m} u_{\gamma,1} \right.
  \\
  + \left. \overline{d^c}'_{\!\!\alpha,l} \left( U^{d\top}_R
    \right)_{l1} \overline{u^c}_{\!\beta,1} \left( u_{\gamma,1} \left(
        U^e_L \right)_{1j} e'_j - d'_{\gamma,m} \left( U^d_L
      \right)_{1m} \left(U^{\nu}_L \right)_{1j} \nu'_j \right) \right]
  + \text{h.c.}
\end{multline}

Let us analyze the mixing pattern for the down quarks in the orbifold
model.  Since the mass matrices of down quarks and charged leptons
both have the form Eq.~(\ref{eq:matrix-structure}), $m^d \sim m^e \sim
m$, $U_R^d$ and $U_L^e$ have the same structure but, in general,
coefficients ${\cal O}(1)$ will be different.  For $\mu_1,\mu_2\ll
\mu_3$, we obtain for the right-handed down-type quarks and
left-handed charged leptons,
\begin{align} \label{eq:down-right}
  U^d_R \sim U^e_L \sim V_4 V_3 = \left( \begin{array}{cccc}
      -\frac{\tM_2}{\tM_{12}} & \frac{\tM_1 \left( \tm_3
          \tM_3-\mu_3\tM_4 \right)}{\bar\mu_3\, \tM\, \tM_{12}} &
      -\frac{\tM_1 \left( \tm_3 \tM_4+\mu_3 \tM_3
        \right)}{\bar\mu_3\, \tM^2} & \frac{\tM_1}{\tM}
      \\
      \rule[-5mm]{0mm}{13mm} \frac{\tM_1}{\tM_{12}} & \frac{\tM_2
        \left( \tm_3 \tM_3-\mu_3 \tM_4 \right)}{\bar\mu_3\,
        \tM\, \tM_{12}} & -\frac{\tM_2 \left( \tm_3 \tM_4+\mu_3
          \tM_3 \right)}{\bar\mu_3\, \tM^2} & \frac{\tM_2}{\tM}
      \\
      \rule[-6mm]{0mm}{13mm} 0 & -\frac{\tm_3}{\bar\mu_3}
      \frac{\tM_{12}}{\tM} & -\frac{\tm_3 \tM_3\tM_4-\mu_3
        \left( \tM_1^2+\tM_2^2+\tM_4^2 \right)}{\bar\mu_3\, \tM^2} &
      \frac{\tM_3}{\tM}
      \\
      \rule[0mm]{0mm}{5mm} 0 & \frac{\mu_3}{\bar\mu_3}
      \frac{\tM_{12}}{\tM} & \frac{\tm_3 \left(
          \tM_1^2+\tM_2^2+\tM_3^2 \right)-\mu_3
        \tM_3\tM_4}{\bar\mu_3\, \tM^2} & \frac{\tM_4}{\tM}
    \end{array} \right) ,
\end{align}
up to the two-dimensional mixing matrix for the second and third
generation discussed in Section~2. 

The rotation matrices of the left-handed down quarks and right-handed
leptons are obtained by diagonalizing $m^{\top}m$, which leads to
small mixing angles, with
\begin{align} \label{eq:relations}
  U^d_L & = V_\ckm\, \sim U_R^e \ .
\end{align}
The unitary matrices $U^e_L$, $U^d_R$, $U^d_L$ and $U^e_R$ determine
the coefficients of the proton decay operators in
Eq.~(\ref{eff6D-VU}).

To make a comparison with ordinary 4D GUT models, we consider the
flavour structure of two {\sf SU(5)} models described in
Refs.~\cite{sy99,Altarelli02}. These models make use of the
Froggatt-Nielsen mechanism \cite{froggatt} where a global ${\sf
  U(1)}_F$ flavour symmetry is broken spontaneously by the VEV of
gauge singlet field $\Phi$ at a high scale. Then the Yukawa couplings
arise from the non-renormalizable operators,
\begin{align}
  h_{ij} = g_{ij} \left( \frac{\VEV{\Phi}}{\Lambda} \right)^{Q_i+Q_j}
  \ .
\end{align}
Here, $g_{ij}$ are couplings $\mathcal{O} \left( 1 \right)$ and $Q_i$
are the ${\sf U(1)}_F$ charges of the various fermions.  Particularly
interesting is the case with a `lopsided' family structure, where the
chiral charges are different for ${\bf 5}^*$ and ${\bf 10}$ of the
same family.  The two examples \cite{sy99,Altarelli02} with
phenomenologically allowed lopsided charges are given in
Table~\ref{tb:su5-charges}.

Note that in these models the large neutrino mixing is explained by a
large mixing of ${\bf 5^*}$-plets, which is analogous to the large
mixing of lepton doublets and right-handed down quarks in the 6D model
described above.  Contrary to the 6D model, this does not determine
the ${\sf U(1)}_F$ charges of the right-handed neutrinos. For proper
choices these models also lead to successful baryogenesis via
leptogenesis \cite{by99}.

%-----------------Table: SU(5)-charges-----------------------------
\begin{table}
  \centering
  \begin{tabular}{c|ccc|ccc}
    $Q_F$ & 
    $\mathbf{10}_3$ & $\mathbf{10}_2$ & $\mathbf{10}_1$  & 
    $\mathbf{5}^*_3$ & $\mathbf{5}^*_2$ & $\mathbf{5}^*_1$ \\
    \hline
    model A & 0 & 1 & 2 & $a$ & $a$ & $a+1$ \\
    model B & 0 & 3 & 5 & 0 & 0 & 2 
  \end{tabular}
  \caption{{\sl ${\sf U(1)}_F$ charges of the {\sf SU(5)} fields;
      $a=0,1$}. 
    \label{tb:su5-charges}}
\end{table}
%-----------------Ende Table----------------------------------------

The charge assignments determine the structure of the Yukawa matrices.
In model A \cite{sy99}, corresponding to the semi-anarchical model of
\cite{Altarelli02}, the couplings for down quarks and charged leptons
read
\begin{align} 
  h_d \sim h_e \sim \epsilon^a \left(
    \begin{array}{lll}
      \epsilon^{3} & \epsilon^{2} & \epsilon^{2} \\ 
      \epsilon^{2} & \epsilon  & \epsilon  \\ 
      \epsilon  & 1 & 1
    \end{array}
  \right) \ ,
\end{align}
where the parameter $\epsilon = \VEV{\Phi}/\Lambda \sim 1/17$ controls
the flavour mixing.  Diagonalisation of the Yukawa matrices yields
\begin{align} \label{eq:lopsided-mixing}
  U^d_L = V_\text{\sc ckm} \sim U_R^e & \sim \left( 
    \begin{array}{lll}
      1 & \epsilon  & \epsilon^{2} \\ 
      \epsilon  & 1 & \epsilon  \\ 
      \epsilon^{2} & \epsilon  & 1
    \end{array}
  \right) \ , &%\\
  U^d_R \sim U_L^e  & \sim \left( 
    \begin{array}{lll}
      1 & \epsilon  & \epsilon \\ 
      \epsilon  & 1 & 1  \\ 
      \epsilon & 1  & 1
    \end{array}
  \right) \ .
\end{align}
In model B, which corresponds to the hierarchical $H_\mathit{II}$
model of \cite{Altarelli02}, the structure is similar, while the small
parameter is instead $\lambda \sim 0.35 $, such that $\lambda^2 \sim
\epsilon$.

\subsection{Decay rates and branching ratios}

To calculate the decay rates, we have to evaluate the hadron matrix
elements $\langle \text{\sl PS} \left| {\cal O} \right| p \rangle$,
which describe the transition from the proton via the three-quark
operator ${\cal O}$ to a pseudo scalar meson.  The various matrix
elements are calculated from the basic element
\begin{align}
  \alpha\, P_L\, u_p & = \epsilon_{\alpha\beta\gamma} \left\langle 0
    \left| \left( d_R^\alpha\, u_R^\beta \right) u_L^\gamma \right| p
  \right\rangle
\end{align}
with the aid of chiral perturbation theory \cite{claudson,chadha};
$u_p$ denotes the proton spinor.  The decay rates for the different
channels are given in Table~\ref{tb:chiral}.  Here $m_p$, $m_\pi$,
$m_K$ and $m_\eta$ denote the masses of proton, pion, kaon and eta,
respectively, and $f_\pi$ is the pion decay constant;
$m_B=1.15\,\text{GeV}$ is an average baryon mass according to
contributions from diagrams with virtual $\Sigma$ and $\Lambda$;
$D=0.80$ and $F=0.46$ are the symmetric and antisymmetric {\sf SU(3)}
reduced matrix elements for the axial-vector current \cite{cabibbo03}.
The matrix element $\alpha$ is evaluated by means of lattice QCD
simulations; its absolute value varies in the range
$\left(0.003-0.03\right)\text{GeV}^3$.  We will choose
$\left|\alpha\right|=0.01\,\text{GeV}^3$ (see the recent discussion in
Ref.~\cite{jlqcd-cppacs}).

%---------------------------Table--------------------------------------
\begin{table}
  \begin{align*}
    \Gamma(p\to e_j^+ \pi^0) & = \frac{(m_p^2-m_{\pi^0}^2)^2}{32\pi
      m_p^3 f_\pi^2}\, {\alpha^2}\, A^2\; G_G^2 \left(
      \frac{1+D+F}{\sqrt{2}} \right)^2\,C_{udue_j}^2
    \\
    \Gamma(p\to \bar\nu_j \pi^+) & =
    \frac{(m_p^2-m_{\pi^\pm}^2)^2}{32\pi m_p^3 f_\pi^2}\,
    {\alpha^2}\, A^2\; G_G^2 \left( 1+D+F \right)^2\, C_{udd\nu_j}^2
    \\
    \Gamma(p\to e_j^+ K^0) & = \frac{(m_p^2-m_{K^0}^2)^2}{32\pi
      m_p^3 f_\pi^2}\, {\alpha^2}\, A^2\; G_G^2 \left(
      1+(D-F)\frac{m_p}{m_B} \right)^2\, C_{usue_j}^2
    \\
    \Gamma(p\to \bar\nu_j K^+) & =
    \frac{(m_p^2-m_{K^\pm}^2)^2}{32\pi m_p^3 f_\pi^2}\, {\alpha^2}
    A^2\; G_G^2 \left[ \left( \frac{2}{3}D \frac{m_p}{m_B} \right)
      C_{usd\nu_j} \!+\! \left( 1+\frac{D+3F}{3}\frac{m_p}{m_B}
      \right) C_{uds\nu_j} \right]^2
    \\
    \Gamma(p\to e_j^+ \eta) & = \frac{(m_p^2-m_\eta^2)^2}{32\pi
      m_p^3 f_\pi^2}\, {\alpha^2}\, A^2\; G_G^2 \left(
      \frac{1+D-3F}{\sqrt{6}} \right)^2\,C_{udue_j}^2
  \end{align*}%} 
  \caption{Partial widths of proton decay channels \cite{jlqcd}.
    \label{tb:chiral}}
\end{table}
%------------------------End Table-------------------------------------

In the decay rates listed in Table~\ref{tb:chiral} the coupling
constant $G_G=g_5^2/(M_{\cal X}^\text{eff})^2$ is given by
Eqs.~(\ref{can}), (\ref{eq:sum-der}) and (\ref{eq:der-int}).  The
operators have to be evolved from the GUT scale down to the hadronic
scale, which is described by the factor $A=A^{\text{SD}}\cdot
A^{\text{LD}}$. It contains both a short-distance contribution
$A^{\text{SD}}=2.37$, for the evolution from the GUT scale to the
SUSY-breaking scale, and a long-distance contribution
$A^{\text{LD}}=1.43$, for the evolution from the SUSY-breaking scale
to 1\,GeV \cite{cw03}. The effect of lepton masses is neglected.

The quark and lepton mixing patterns discussed above fix the
coefficients $C^2_{ijkm}$ in the decay rates. As an example, for the
process $p\to e^+\pi^0$, we obtain from Eq.~(\ref{eff6D-VU}),
\begin{align}
  C^2_{udue} = 4 \left[ \left( U^{e}_R \right)_{11} \left( U^d_L
    \right)_{11} \right]^2 + \left[ \left( U^{d}_R \right)_{11} \left(
      U^e_L \right)_{11} \right]^2 \ .
\end{align}
The coefficients for the other processes can be read off analogously.
The decay rates in Table~\ref{tb:chiral} have the same form as the
decay rates determined by dimension-5 operators. The difference lies
in the coefficients $C_{ijkm}$ and the coupling constant $G_G$.

We now start with the simplest case of our orbifold model, with
$U^d_R=U_L^e$, and with degenerate masses $\tM$ and
$\tm_3=\mu_3$, which we denote as case I.  The mixing matrix for
right-handed down quarks and left-handed charged leptons,
Eq.~(\ref{eq:down-right}), is then simply given by
\begin{align} \label{eq:case1}
  U^d_R = U^e_L = \left( \begin{array}{cccc}
      -\frac{1}{\sqrt{2}} & 0 & -\frac{1}{2} & \frac{1}{2} \\
      \rule[-5mm]{0mm}{13mm}
      \frac{1}{\sqrt{2}} & 0 & -\frac{1}{2} & \frac{1}{2} \\
      \rule[-5mm]{0mm}{10mm}
      0 & -\frac{1}{\sqrt{2}} & \frac{1}{2} & \frac{1}{2} \\
      \rule{0mm}{5mm} 0 & \frac{1}{\sqrt{2}} & \frac{1}{2} &
      \frac{1}{2}
    \end{array} \right) \ ,
\end{align}
thus the state $d^c_1$ has no strange-component. For the
current-current operators we then obtain from Eq.~(\ref{eff6D-VU}),
\begin{multline} \label{eq:lagr-caseI}
  {\cal L}_\text{eff} \simeq \frac{g_5^2}{(M_{\cal X}^\text{eff})^2}
  \epsilon_{\alpha\beta\gamma} \left[ 2\,V_{ud}^2\, \overline{e^c}\,
    \overline{u^c_\alpha}\, d_\beta\, u_\gamma +\frac{1}{2}\,
    \overline{d^c_\alpha}\, \overline{u^c_\beta}\, u_\gamma\, e
    +2\,V_{ud} V_{us}\, \overline{\mu^c}\, \overline{u^c_\alpha}\,
    d_\beta\, u_\gamma \right.
  \\[2mm]
  +2\,V_{ud} V_{us}\, \overline{e^c}\, \overline{u^c_\alpha}\,
  s_\beta\, u_\gamma + 2\,V_{us}^2\, \overline{\mu^c}\,
  \overline{u^c_\alpha}\, s_\beta\, u_\gamma
  \\[2mm]
  - \left. \sum_{j=1}^3 \frac{1}{\sqrt{2}}\, \left( U^\nu_L
    \right)_{1j}\, \overline{u^c_\alpha}\, \overline{d^c_\beta}\,
    \left\{ V_{ud}\, d_\gamma\, + V_{us}\, s_\gamma
      \vphantom{\frac12}\right\}\, \nu_j \right] + \text{h.c.} \ ,
\end{multline}
where the fermions are now mass eigenstates.  From this equation we
can read off the coefficients of the various coefficients $C_{ijkm}$
appearing in the decay rates (cf. Table~\ref{tb:chiral}),
\begin{align} \label{eq:so10-coefficients-case1}
    C_{udue}^2 & = 4\,V_{ud}^4 + \frac{1}{4} \ , &
    C_{usu\mu}^2 & = 4\,V_{us}^4 \ , &
    C_{udu\mu}^2 & = C_{usue}^2 = 4\,V_{us}^2 V_{ud}^2 \ , 
    \nonumber \\
    C_{udd\nu}^2 & = \frac{1}{2}\,V_{ud}^2 \ , &
    C_{uds\nu} & = \frac{1}{\sqrt{2}}\,V_{us} \ , &
    C_{usd\nu} & = 0 \ ,
\end{align}
where we have used $\sum_{j=1}^3 \left( U^\nu_L\right)_{1j} \left(
  U^\nu_L \right)_{1j}^* =1$.

The numerical results for the branching ratios are listed in
Table~\ref{tb:result}. Note that the effect of the derivative
operators is negligible for \mbox{$c_5=c_6=1$}. For the listed
branching ratios the corrections are less than 3\%.

%---------------------------Table--------------------------------------
\begin{table}[t]
  \centering
  \begin{tabular}{c|cc|c}
    decay channel & \multicolumn{3}{c}{Branching Ratios [\%]} \\
    \cline{2-4}
    & \multicolumn{2}{c|}{6D {\sf SO(10)}} &
    ${\sf SU(5)}\times {\sf U(1)}_F$ \\ 
    & case I & case II &  models A \& B\\ 
    \hline
    $e^+\pi^0$ & 75 & 71 & 54 \\
    $\mu^+\pi^0$ & 4 & 5 & <\,1 \\
    $\bar\nu\pi^+$ & 19 & 23 & 27 \\
    $e^+ K^0$ & 1 & 1 & <\,1 \\
    $\mu^+ K^0$ & <\,1 & <\,1 & 18 \\
    $\bar\nu K^+$ & <\,1 & <\,1 & <\,1 \\
    $e^+\eta$ & <\,1 & <\,1 & <\,1 \\
    $\mu^+\eta$ & <\,1 & <\,1 & <\,1 \\
  \end{tabular}
  \caption{Resulting branching ratios and comparision with \mbox{${\sf
        SU(5)}\times {\sf U(1)}_F$}.  \label{tb:result}} 
\end{table}
%------------------------End Table-------------------------------------

To compare the branching ratios of the 6D model with those of the two
4D GUT models described above, we assume that some mechanism
suppresses or avoids the proton decay arising from dimension-5
operators. The coefficients $C^2_{ijkl}$ can then be derived from
Eq.~(\ref{eff4D-VU}) using the mixing matrices given in
Eq.~(\ref{eq:lopsided-mixing}). For model A, they read
\pagebreak
\begin{align} \label{eq:su5-coefficients-A}
  C^{2}_{udue} & \simeq C^{2}_{usu\mu} \simeq 4 \ , &
  C^{2}_{udu\mu} & \simeq C^{2}_{usue} \simeq 4\,\epsilon^{2}\ , 
  \nonumber \\
  C^{2}_{udd\nu} & \simeq \frac{1}{2} \ , &
  C_{uds\nu} & \simeq \frac{1}{\sqrt{2}} \ , & 
  C_{usd\nu} & \simeq 2\,\epsilon \ .
\end{align}
In model B, $\epsilon$ is replaced by $\lambda^2$.  The differences
between the branching ratios of the two models are not significant.

The difference between the 6D {\sf SO(10)} and the 4D {\sf SU(5)}
models is most noticeable in the channel $p\to \mu^+ K^0$. This is due
to the absence of second and third generation weak eigenstates in the
current-current operators and the vanishing (12)-component in $U_R^d$
and $U_L^e$ in the case of the 6D model.  Hence, the decay $p\to \mu^+
K^0$ is doubly Cabibbo suppressed.  This effect is a direct
consequence of the localization of the `first generation' to the
Georgi-Glashow brane.

Let us now consider the general case, where the $\tM^{(d,e)}$ are not
degenerate, and where $\mu_3$ and $\tm^{(d,e)}_3$ differ as well.
From Eq.~(\ref{eq:down-right}) we see that the strange component in
$d^c_1$ does not vanish anymore, but it is smaller than the bottom
component.  We have studied several cases whose results agree
remarkably well.  As an illustration, consider the case where
$\tm^d_3=2\mu_3$ and $\tm^e_3=3\mu_3$, with non-degenerate heavy
masses $\tM^{d}_1 : \tM^{d}_2 : \tM^{d}_3 : \tM^{d}_4 =
\frac{1}{2}:\frac{1}{\sqrt{2}}:\frac{1}{\sqrt{2}}:1$ and $\tM^{e}_1 :
\tM^{e}_2 : \tM^{e}_3 : \tM^{e}_4 =
\frac{1}{2}:\frac{1}{\sqrt{2}}:1:\frac{1}{2}$ (case II).  The
branching ratios are listed in Table~\ref{tb:result}; the differences
between the two cases are indeed small.

The most striking difference is the decay channel $p\to \mu^+ K^0$,
which is suppressed by about two orders of magnitude in the 6D model
with respect to 4D models.  It is therefore important to determine an
upper limit for this channel in the 6D model.  Varying the mass
parameters in the range $\widetilde{M}_j/\widetilde{M}=0.1 - 1$ and
$\tm_3^{d,e}/\mu_3=0.1 - 10$, we find
\begin{align}
  \frac{\Gamma(p\to \mu^+ K^0)}{\Gamma(p\to e^+ \pi^0)} \lesssim 5\,\%
  \ .
\end{align}

Finally, a limit on the compactification scale can be derived from the
decay width of the dominant channel $p\to e^+ \pi^0$.  Neglecting the
suppressed contributions from the derivative operators, we obtain the
analytic expression
\begin{align}
  \Gamma(p\to e^+ \pi^0) & \simeq \frac{(m_p^2-m_{\pi^0}^2)^2}{32\pi
    m_p^3 f_\pi^2}\ {\alpha^2} A^2\ \left(
    \frac{1+D+F}{\sqrt{2}} \right)^2 \\[2mm]
  & \quad \times\ \frac{\pi^2}{16\,M_c^4} \left(\ln \left(
      \frac{M_*}{M_c} \right) + 2.3 \right)^2 \left[\, 4 V_{ud}^4 +
    \frac{\tM_2^{d\,2}}{\tM_1^{d\,2}+\tM_2^{d\,2}}
    \frac{\tM_2^{e\,2}}{\tM_1^{e\,2}+\tM_2^{e\,2}} \right] \NO\\[2mm]
  &\simeq \(\frac{9\times 10^{15}\ \text{GeV}}{M_c}\)^4
  \(\frac{\alpha}{0.01\ \text{GeV}^3}\)^2 \(5.3 \times 10^{33}\ 
  \text{yrs}\)^{-1}\;.
\end{align}
With $M_*=10^{17}\,\text{GeV}$ and $\tM^{d,e}_{1,2}={\cal
  O}\left(1\right)$, the experimental limit $\tau\geq 5.3\times
10^{33}$ yields $M_c \geq M_c^\text{min} \simeq 9\times
10^{15}\,\text{GeV}$, which is very close to the 4D GUT scale.  The
lower bound corresponds to $M_*/M_c^\text{min} = 12$ and $(M_{\cal
  \chi}^\text{eff})^2/(M_c^\text{min})^2 = 3.72$.

\section{Conclusions \label{se:concl}} 

We have studied proton decay in a 6D {\sf SO(10)} orbifold GUT model,
which is determined by dimension-6 operators.  The characteristic
features of the model are the different breakings of the {\sf SO(10)}
symmetry at different points in the extra dimensions and the
associated localization of some quarks and leptons.  We find that,
like in 5D orbifold GUTs \cite{hm02}, the proton decay rate is
enhanced and the branching ratios are strongly affected by the
quark-lepton `geography'.
 
The summation over Kaluza-Klein towers depends logarithmically on the
cutoff scale $M_*$ in 6D, contrary to 5D, where the sum is finite.
Identifying the cutoff with the 6D Planck mass, i.e.  $M_* \simeq
10^{17}$~GeV, the SuperKamiokande bound on the proton lifetime leads
to the lower bound on the compactification scale $M_c > 9\times
10^{15}$~GeV.  On the other hand, the approximate unification of gauge
couplings suggests $M_c \simeq \L_\text{GUT} \simeq 2\times
10^{16}$~GeV.  This yields the proton lifetime $\t(p\rightarrow
e^+\p^0) \simeq 1\times 10^{35}$~yrs which, remarkably, lies within
the reach of the next generation of large volume detectors!

The peculiar flavour structure of our orbifold GUT model leads to
characteristic signatures in the branching ratios of proton decay, in
particular the strong suppression of the mode $p \rightarrow \mu^+
K^0$ compared to the predictions of 4D models.  The reason is the
higher-dimensional quark-lepton `geography' and the related
non-universal couplings of GUT bosons to fermions.  Such a pattern can
be tested already with a handful of events!  Hence, the discovery of
proton decay may not only confirm the most striking prediction of
grand unification, but it might also reveal its higher
dimensional origin.

\bigskip  

\noindent 
We would like to thank A.~Hebecker for helpful discussions.  L.C.
would also like to thank J.~Ellis, and A.~Romanino, C.~Scrucca and the
other participants of the CERN Phen Club on 26/2/2004 for useful
discussions and suggestions.  The work of D.E.C. was supported by
Fundação para a Ciência e a Tecnologia under the grant
SFRH/BPD/1598/2000.

\end{document}